\documentstyle[12pt]{article}
\newcommand{\bea}{\begin{eqnarray}}\newcommand{\eea}{\end{eqnarray}}
\def\spmapright#1{\smash{\mathop{\hbox to 1.3cm{\rightarrowfill}}
        \limits^{#1}}}
\def\sbmapright#1{\smash{\mathop{\hbox to 1.3cm{\rightarrowfill}}
        \limits^{#1}}}

\begin{document}
\begin{titlepage}
\nopagebreak
\begin{flushright}
UT-667 \\
February 1994\\ hep-th/xxxxxxx \\ ~
\end{flushright}
\begin{center}
{\LARGE
Global structure and thermodynamic property of \\
the 4-dimensional twisted Kerr solution}
\vfill
{\large Tadashi Okai}\\
\vspace{1cm}
\it
Department of Physics, University of Tokyo\\
Bunkyo-ku, Tokyo 113, Japan \\
\rm
e-mail: okai@tkyux.phys.s.u-tokyo.ac.jp \\
{}~~~~~~~~~~okai@tkyvax.phys.s.u-tokyo.ac.jp \\
{}~~~~~~~~~~~okai@danjuro.phys.s.u-tokyo.ac.jp
\\
\end{center}
\vfill
\begin{abstract}
Rotating stringy black hole solutions with non-vanishing dilaton
$\phi$,
antisymmetric tensor $B_{\mu\nu}$, and $U(1)$ gauge field
$A_{\mu}$ are investigated.
Both Boyer-Lindquist-like and Kerr-Schild-like coordinate are
constructed. The latter is utilised to
construct the analytically extended spacetime.
The global structure of the resulting extended spacetime is almost
identical to that of the Kerr.
In carrying out the analytic extension,
the radial coordinate should be
suitably chosen so that we can avoid singularity caused by the twisting.
The thermodynamic property of the stringy black hole is examined
through the injection of test bodies into the black hole.
It is shown that one cannot change a black hole configuration
into a naked singularity by way of throwing test bodies into
the black hole.
The global $O(2,3)$ symmetry and the preservation of the
asymptotic flatness are discussed.
When we impose stationarity, axisymmetry, and asymptotic flatness,
there is no other twisting than the one pointed out by
A.Sen\cite{sen}. All the other elements of $O(2,3)$ either break
the asymptotic flatness, or cause only coordinate transformations
and gives no physical change.
\end{abstract}
\vfill
\end{titlepage}
%
\renewcommand{\theequation}{\thesection.\arabic{equation}}
\setcounter{equation}{0}
\section{Introduction}
Exact solutions in General Relativity have  given us simplified,
but fairly qualitative features of spacetimes. Theoretical
developments of black holes owe exact solutions for their
establishment.
Solving the Einstein equation exactly, however, is very difficult
and only when we impose some symmetry, can we solve it\cite{kramer}.
One of the largest symmetry we often consider
is to impose staticity and spherical symmetry. Then
the metric and matter fields become functions of one variable of the
radial coordinate
and the equations of motion become a suit of ordinary
differential equations. Some can be solved analytically
(and some numerically)
to get physically interesting solutions. When one loosens
the spherical symmetry down to axisymmetry, the difficulty increases
a great deal and direct quadrature is impossible in general.
Several remarkable techniques (which include the introduction of
potentials\cite{ernst},
superposition of known solutions\cite{superposition},
or transformations of known solutions to new ones\cite{hkx})
have been devised to circumvent this
difficulty. These techniques are unquestionably important for solving
the Einstein equation.

On the other hand, the intensive research of string theories in
1980's\cite{gsw}
has enriched the models including gravity\cite{string}.
Characteristic feature of
these models, inspired by string theory or Kaluza-Klein theory, is
inclusion of the dilaton.
The investigation has revealed that these models allow black hole
solutions \cite{garf}\cite{maeda}\cite{limit}\cite{ichinose}
and that there exist transformations \cite{leipzig}\cite{sen}
that generate
new solutions from old ones in these models, just as well as
one can obtain charged solutions
from uncharged ones through the Harrison transformation in
Einstein-Maxwell theory\cite{harrison}.
We know from this that the low energy heterotic string theory
possesses a global $O(d,d+p)$ symmetry when the spacetime admits
$d$ Killing vectors and the theory contains $p$ $U(1)-$ gauge
fields\cite{sen}.

This paper is motivated
\it
1) to clarify the global structure and thermodynamic property
\rm
of the rotating charged black hole solution\cite{sen} inspired
by the heterotic string theory, and
\it
2) to examine the possibility of producing other black hole solutions
by $O(2,3)$ ($d=2$ and $p=1$ of $(d,d+p)$) transformation.
\rm
As we will show in section 4, analytic extension of the twisted Kerr
solution is obtained by introducing a new radial coordinate
$R:=\sqrt{r^2 +m (\cosh\alpha -1)r}$, where $\alpha$ is a twist
parameter. The global structure and thermodynamic property are
almost identical to those of the Kerr.
We will show in section 6 that under fairly general conditions,
there is only one possibility of the twisting that preserves the
asymptotic flatness. That is,
$\Omega_{35}:=\exp (\alpha(e_{35}+e_{53})) \in O(2,3)$
is the only one, which was discussed by A.Sen\cite{sen}.

The organisation of this article is the following.
In section 2, we review the result of A.Sen et al.\cite{sen}
that the action (\ref{haction})
allows the global $O(2,3)$ symmetry when one assumes that the
spacetime is stationary and axisymmetric.
(We consider only the case of $d=2$ and $p=1$.)
The twisted Kerr solution is
obtained by the action of $\Omega_{35}$
on the Kerr solution.

In section 3, we construct a Boyer-Lindquist-like coordinate and
consider the geodesic equations on the twisted Kerr metric background.

A Kerr-Schild-like coordinate is constructed and the
global structure of the spacetime is obtained from the analytically
extended solution. This is written in section 4.

Section 5 treats the gedanken experiments of throwing test
particles\cite{wald.ann} into the twisted Kerr solution.
We will show that one cannot decrease
the area of the horizon by any means of throwing test bodies.
Moreover, the condition in which test bodies fall inside the horizon
exactly reproduces the thermal structure of the twisted Kerr
solution\cite{beken}\cite{car}.

In section 6, we discuss the global $O(2,3)$ transformation
in view of the asymptotic flatness of the spacetime.
Starting with a known solution, we can obtain
several solutions by acting elements of $O(2,3)$.
However, some elements of $O(2,3)$ break the asymptotic flatness.
We will investigate the infinitesimal action of $o(2,3)$ and show
that $\Omega_{35}$ is the only
transformation that preserves the asymptotic flatness.

Section 7 is devoted to summary and conclusion.
%
\setcounter{equation}{0}
\section{Global $O(2,3)$ symmetry}
In this section, we review the construction of a rotating charged
stringy black hole solution by \lq\lq twisting" the Kerr solution.
We consider the following action, which includes
the metric $g_{\mu\nu}$, $U(1)-$ gauge field $A_{\mu}$, the dilaton
field $\phi$, and
the antisymmetric
tensor field $B_{\mu\nu}$:
\bea
S=\int  d^{4}x \sqrt{-g}\left( R-2(\partial\phi )^{2}-e^{-2\phi}F^{2}
-\frac{1}{12}e^{-4\phi }H^{2}\right) ,
\label{haction}
\eea
where we put
\bea
H_{\alpha\beta\gamma}:=\partial _{\alpha}B_{\beta\gamma}
+\partial _{\beta}B_{\gamma\alpha}+\partial _{\gamma}B_{\alpha\beta}
-2(A_{\alpha}F_{\beta\gamma}+A_{\beta}F_{\gamma\alpha}
+A_{\gamma}F_{\alpha\beta}). \nonumber
\eea
It is pointed out by A.Sen et al. that if there are $d$ independent
Killing vector fields, the action (\ref{haction}) has global
$O(d,d+1)$ symmetry\cite{sen}.

To make this symmetry manifest, we temporarily redefine
the metric $g_{\mu\nu}$ of
eq.(\ref{haction}) by $e^{2\phi}g_{\mu\nu}$,
and suitably rescale $\phi$ and $A_{\mu}$ by
some constant factors. Then
the action (\ref{haction}) is equivalent to
\bea
S=\int d^{4}x\sqrt{-g}
e^{-\phi }\left( R+(\partial\phi)^{2}-\frac{1}{8}F^{2}
-\frac{1}{12}H^{2}\right) .
\label{varhact}
\eea
The $H_{\alpha\beta\gamma}$ is now defined by
\bea
H_{\alpha\beta\gamma}:=\partial _{\alpha}B_{\beta\gamma}
+\partial _{\beta}B_{\gamma\alpha}+\partial _{\gamma}B_{\alpha\beta}
-\frac{1}{4}\left( A_{\alpha}F_{\beta\gamma}+A_{\beta}F_{\gamma\alpha}
+A_{\gamma}F_{\alpha\beta}\right) . \nonumber
\eea
When one treats stationary and axisymmetric solutions, which we will
do throughout this paper, they have global $O(2,3)$ symmetry.
We will use eq.(\ref{varhact}) only for the purpose of showing
the global $O(2,3)$ symmetry. All physical discussions will be
made from the viewpoint of the action (\ref{haction}) throughout
this paper.

Let us first recall the $O(2,3)$ symmetry that generates new solutions
from a known one. We assume that the system be a
stationary and axisymmetric rotating body. That is, the system has
a timelike (at least outside the horizon) coordinate $t$ and spacelike
periodic coordinate $\varphi$ such that
$\displaystyle
\left( \frac{\partial}{\partial t} \right)$ and $\displaystyle
\left( \frac{\partial}{\partial \varphi} \right) $
are Killing vectors. The \lq\lq rotating body" means the system is
invariant under the simultaneous inversion of the signs of
$(t, \varphi)$. (see, for example, ref.\cite{chandra}.)
The metric is then written as
\bea
g_{\mu\nu}=\left(
\matrix{\hat{g}_{mn}&0\cr 0&\tilde{g}_{\alpha\beta}\cr}
\right), \nonumber
\eea
where both $\hat{g}_{mn}$ and $\tilde{g}_{\alpha \beta }$ are
functions of $(r, \theta)$ and independent
of $(t, \varphi)$. They are respectively 2 dimensional metrics of
subspaces spanned by $(t, \varphi)$ and $(r, \theta)$. Hereafter we
will
use the symbols of $~~\hat{}~~$ and $~~\tilde{}~~$
as the $(t, \varphi)-$ and $(r, \theta)-$
components of tensor fields. Respecting the symmetry of
the system, the configuration of the other fields is in the following
form:
\bea
\phi =\phi (r,\theta ), \;\;\;
B_{\mu\nu}=\left( \matrix{\hat{B}_{mn}&0\cr 0&\tilde{B}_{\alpha\beta}
\cr } \right), \;\;\;
A_{\mu }=\left( \matrix{\hat{A}_{m}\cr  \tilde{A}_{\alpha } \cr }
\right). \nonumber
\eea
Again, these are functions of $(r, \theta)$ only and independent of
$(t, \varphi)$.

Some calculations are needed to see the manifest global $O(2,3)$
symmetry. They are all results of straightforward calculations. We
show them here in the form of a proposition:\\
\\
{\bf Proposition} (A.Sen\cite{sen})\\
\it
The identity 1), 2), 3), and 4) hold.\\
$\displaystyle
\mbox{1)}~~~~~e^{-\phi}\sqrt{-g}
(R+(\partial _{\mu} \phi )(\partial _{\nu } \phi )g^{\mu\nu })$\\
$\displaystyle
{}~~~~~~~~~~=e^{-\phi}\sqrt{-g}(R^{(2)}
+\frac{1}{4}\mbox{\rm tr}(\partial_{\alpha}\hat{g}
\cdot\partial_{\beta}\hat{g}^{-1})
\tilde {g}^{\alpha \beta }
+\partial _{\alpha }(\phi -\log  \sqrt{-\hat{g}})
\cdot  \partial _{\beta }(\phi -\log  \sqrt{-\hat{g}})
\tilde {g}^{\alpha \beta })$\\
$\displaystyle
{}~~~~~~~~~~~~~+\mbox{\rm total divergence}$,\\
where $R^{(2)}$ is the scalar curvature of the 2
dimensional subspace
spanned by $(r, \theta)$, and the trace is taken over
the $(t, \varphi)-$ space.\\
$\displaystyle
\mbox{2)}~~~~~
\sqrt{-g}e^{-\phi }=\sqrt{-\hat{g}}\sqrt{\tilde {g}}e^{-\phi }
=\sqrt{\tilde {g}}e^{-\chi }, \;\;\;\mbox{where we put}\;\; \chi
:=\phi -\log \sqrt{-\hat{g}}. \;\;
$\\
$\displaystyle
\mbox{3)}~~~~~
-\frac{1}{12}H^{2}-\frac{1}{8}F^{2}
=\frac{1}{4}\mbox{\rm tr}
(H_{\alpha }\hat{g}^{-1}H_{\beta }\hat{g}^{-1})
\tilde {g}^{\alpha \beta }
-\frac{1}{4}\mbox{\rm tr}
(\partial _{\alpha }\hat{A}\hat{g}^{-1}\partial _{\beta }
\hat{A}')\tilde {g}^{\alpha \beta }
-\frac{1}{12}\tilde {H}^{2}-\frac{1}{8}\tilde {F}^{2}.
$\\
Here we denote the transposition of a matrix $M$ by $M':={}^{t}\!M$
for simplicity.
We have put the matrix $H_{\alpha}$ as
$(H_{\alpha})_{mn}:=H_{\alpha mn}$. The traces are again taken over
the $(t, \varphi)-$ space.
\\
$\displaystyle
\mbox{4)}~
\left[
\frac{1}{4}\mbox{\rm tr}
(\partial _{\alpha }\hat{g} \partial _{\beta }\hat{g}^{-1})
+\frac{1}{4}\mbox{\rm tr}
(\hat{H}_{\alpha }\hat{g}^{-1}\hat{H}_{\beta }\hat{g}^{-1})
-\frac{1}{4}\mbox{\rm tr}
(\partial_{\alpha}\hat{A}\hat{g}^{-1}\partial_{\beta}\hat{A}')
\right]\tilde {g}^{\alpha \beta }
=\frac{1}{32}\mbox{\rm tr}(\partial _{\alpha }{\cal M}{\cal L}
\partial _{\beta }{\cal M}{\cal L})
\tilde {g}^{\alpha \beta },
$\\
where we defined ${\cal M}$, $k$, $\eta$, and ${\cal L}$ by
\bea
&&{\cal M}:=\left(\matrix{k'-\eta  \cr  k'+\eta  \cr  -\hat{A}' \cr }\right)
\matrix{ \hat{g}^{-1}\cr\cr\cr }
\matrix{( k-\eta & k+\eta  & -\hat{A} )\cr &&\cr&&\cr}=:
\left(\matrix{l&n&u \cr  n'&c&e \cr  u'&e'&f \cr }\right)
\in M_{5}({\bf R}), \nonumber \\
&&k:=-\hat{B}-\hat{g}-\frac{1}{4}\hat{A}\hat{A}',\;\;\;
\eta :=\left(\matrix{-1&0\cr  0&1 \cr }\right), \;\;\;
{\cal L}
     :=\left(\matrix{-1&&&&\cr &1&&&\cr &&1&&\cr &&&-1&\cr &&&&-1\cr}
       \right). \;\; \frac{~~~~~~}{~~~~~~}
\nonumber
\eea
\\
\rm
Using this proposition,
the action (\ref{varhact}) is now rewritten as
\bea
S&=&\int dtd\varphi\int drd\theta\sqrt{\tilde{g}}e^{-\chi}
\left[R^{(2)}
+\partial_{\alpha}\chi\partial_{\beta}\chi\tilde{g}^{\alpha\beta }
-\frac{1}{8}\tilde {F}^{2}
+\frac{1}{32}
   \mbox{tr}(\partial _{\alpha }{\cal M}{\cal L}
   \partial _{\beta }{\cal M}{\cal L})
   \tilde {g}^{\alpha \beta }
\right] \nonumber \\
&&+\mbox{total divergence}.
\label{symact}
\eea
Note that the matrix-valued field ${\cal M}$ is constructed from
$(\hat{g}_{mn}, \hat{B}_{mn}, \hat{A}_{m})$. The
rewritten action (\ref{symact}) is considered as a theory with fields
$(\tilde{g}_{\alpha \beta }, \chi, \tilde{A}_{\alpha }, {\cal M} )$ defined
on the 2 dimensional space $(r, \theta)$. Eq.(\ref{symact}) clearly
possesses the
global $O(2,3)$ symmetry
\bea
{\cal M}\longmapsto \Omega {\cal M}\Omega', \;\;\;\;
\chi\longmapsto\chi \;\;
\left( \;\;
\mbox{ i.e., }
\exp[2\Omega(\phi)]=e^{2\phi}
        \mbox{det}(\Omega(\hat{g}))/\mbox{det}(\hat{g})
        \;\;\right) \nonumber
\eea
with $\Omega$ satisfying the condition
$\Omega'{\cal L}\Omega={\cal L}$. The interpretation between
$(\hat{g}_{ab}, \hat{A}_{a}, \hat{B}_{ab})$ and the
matrix-valued field ${\cal M}$ is given by
\bea
\hat{g}^{-1}&=&\frac{1}{4}\eta (l-n-n'+c)\eta, \nonumber \\
k'&=&2(-l+n)(l-n-n'+c)^{-1}\eta +\eta , \;\;\;
\hat{A}=2\eta  (-l+n)^{-1}u,
\label{m2gab}
\eea
and its inverse is
\bea
l&=&(k'-\eta )\hat{g}^{-1}(k-\eta ), \;\;\;
n=(k'-\eta  )\hat{g}^{-1}(k+\eta ), \nonumber \\
c&=&(k'+\eta  )\hat{g}^{-1}(k+\eta ), \;\;\;
u=(k'-\eta )\hat{g}^{-1}(-\hat{A}), \nonumber \\
e&=&(k'+\eta )\hat{g}^{-1}(-\hat{A}), \;\;\;
f=\hat{A}'\hat{g}^{-1}\hat{A}.
\label{gab2m}
\eea

One immediately recognises that not all the elements of $O(2,3)$
preserve the asymptotic flatness even when one starts with an
asymptotically flat seed metric.
We will consider this problem in section 6.
We act
\bea
\Omega_{35}:=\exp (\alpha(e_{35}+e_{53}))=
\left(
\matrix{1&&&& \cr &1&&& \cr &&\cosh \alpha&&\sinh \alpha \cr
&&&1& \cr &&\sinh \alpha&&\cosh \alpha \cr }
\right) \in O(2,3)
\nonumber
\eea
on the (untwisted) Kerr solution to obtain the twisted Kerr solution.
The untwisted Kerr solution is
\bea
ds^{2}&=&-\left( 1-\frac{2mr}{r^{2}+a^{2}\cos^{2}\theta}\right)dt^{2}
-2\frac{2mra\sin^{2}\theta}{r^{2}+a^{2}\cos^{2}\theta}dtd\varphi
\nonumber \\
&&+\frac{\sin^{2}\theta}{r^{2}+a^{2}\cos^{2}\theta}
[(r^{2}+a^{2})^{2}-(r^{2}-2mr+a^{2}) a^{2}\sin^{2}\theta]d\varphi^{2}
\nonumber \\
&&+\frac{r^{2}+a^{2}\cos^{2}\theta}{r^{2}-2mr+a^{2}}dr^{2}
+(r^{2}+a^{2}\cos^{2}\theta)d\theta^{2},
\nonumber \\
B=A=\phi &=&0.
\nonumber
\eea
The result of the action of $\Omega_{35}$ is given
by
\bea
ds^{2}&=&-\frac{r^{2}+a^{2}\cos^{2}\theta -2mr}
{r^{2}+a^{2}\cos^{2}\theta +\beta  r }dt^{2}
-\frac{2(2m+\beta )ra \sin^{2}\theta }
{r^{2}+a^{2}\cos^{2}\theta +\beta  r }
dtd\varphi \nonumber \\
&&+[(r^{2}+\beta  r+a^{2})^{2}-(r^2 -2mr+a^2 ) a^{2}\sin^{2}\theta ]
\frac{\sin^{2}\theta }
{r^{2}+\beta  r+a^{2}\cos^{2}\theta }d\varphi^{2}
\nonumber \\
&&+\frac{r^{2}+a^{2}\cos^{2}\theta +\beta  r}{r^{2}+a^{2}-2mr}dr^{2}
+(r^{2}+a^{2}\cos^{2}\theta +\beta r)d\theta ^{2}, \nonumber \\
A&=&\frac{-mr\sinh\alpha}{\sqrt{2}(r^{2}+a^{2}\cos^{2}\theta+\beta r)}
(dt-a\sin^{2}\theta d\varphi), \nonumber \\
B_{t\varphi}
&=&\frac{\beta ra\sin^{2}\theta}{r^{2}+a^{2}\cos^{2}\theta +\beta r},
\;\;\;
e^{-2\phi}=
\frac{r^{2}+a^{2}\cos^{2}\theta +\beta r}{r^{2}+a^{2}\cos^{2}\theta}.
\label{senmetric}
\eea

One can easily see that when the seed metric is spherically symmetric
(that is, the Schwarzschild metric), the twisted solution is
\bea
ds^{2}&=&-\left( 1-\frac{2m}{r}\right)
\left( 1+\frac{\beta }{r}\right)^{-1}dt^{2}
+\left( 1+\frac{\beta }{r}\right)
\left( 1-\frac{2m}{r}\right)^{-1}dr^{2}
+r^{2}\left( 1+\frac{\beta }{r}\right) d\Omega,
\nonumber \\
A&=&\frac{-mr\sinh \alpha}{\sqrt{2}(r^{2}+\beta r)}dt, \;\;\;
B_{t\varphi}=0, \;\;\;
e^{-2\phi}=1+\frac{\beta }{r},
\label{twistschwarzschild}
\eea
where we have put $\beta :=(\cosh \alpha -1)m \geq 0$.
One introduces a
new radial
coordinate $\bar{r} :=r+\beta $ to get
\bea
&&ds^{2}=-\left( 1-\frac{2m+\beta }{\bar{r}}\right) dt^{2}
+\left( 1-\frac{2m+\beta }{\bar{r}}\right)^{-1}d\bar{r}^{2}
+\bar{r}^{2}\left( 1-\frac{\beta}{\bar{r}}\right) d\Omega,
\nonumber \\
&&A= -\frac{m\sinh\alpha}{\sqrt{2}\bar{r}}dt, \;\;\; B_{t\varphi}=0,
\;\;\;
e^{2\phi}=1-\frac{\beta}{\bar{r}}. \nonumber
\eea
Global structure and physical properties of this metric are discussed
in detail in ref.s \cite{garf}\cite{maeda}.

What we want to emphasise here is that the twisting by $\Omega_{35}$
produces a new singularity at $r=-\beta$ (or $\bar{r}=0$, see
eq.(\ref{twistschwarzschild})). $r=0$ $(\bar{r}=\beta)$ is a
singularity that has been primordially existing before twisting.
We can roughly say that the twisting $\Omega_{35}$ adds new
singularities in the region $r<0$ of the Schwarzschild metric.
We will see in section 4 that the same phenomenon happens in the case
of the $\Omega_{35}-$ twisted Kerr metric. New singularities do not
appear in $r>0$ and do only in $r\leq 0$ in the twisted Kerr solution
as well as in the twisted Schwarzschild solution.
%
\setcounter{equation}{0}
\section{Boyer-Lindquist coordinate}
The twisted Kerr solution is expressed in a
concise form by \lq\lq diagonalising" the $(t, \varphi)$ subspace of
the metric\cite{globalstr}:
\bea
ds^{2}=-\frac{\Delta }{\rho ^{2}}\left(dt-a\sin^{2}\theta  d\phi \right)^{2}
+\frac{1}{\rho ^{2}}\sin^{2}\theta \left(adt-(R^{2}+a^{2})d\phi \right)^{2}
+\frac{\rho ^{2}}{\Delta }dr^{2}+\rho ^{2}d\theta ^{2},
\nonumber
\eea
where we put
\bea
&&\beta :=m(\cosh \alpha -1), \;\;\;
R^{2}:=r^{2}+\beta r, \nonumber \\
&&\rho ^{2}:=R^{2}+a^{2}\cos^{2}\theta, \;\;\;
\Delta :=r^{2}-2mr+a^{2}. \nonumber
\eea
The electromagnetic field is concisely written as
\bea
A=\frac{Qr}{\rho^{2}}(dt-a\sin^{2}\theta),
\nonumber
\eea
where we put $Q:=-m\sinh \alpha/\sqrt{2}$.
The expression is almost the same as that of the untwisted Kerr
solution except that
some of $r^{2}$'s in the untwisted solution are replaced by
$R^{2}:=r^{2}+\beta r$ in the twisted one. Clearly, the zeros of $g_{tt}$ and
$\Delta$ are the same in both the twisted and the untwisted.
This shows that
the positions of ergosurfaces and horizons remain invariant through
the twisting by $\Omega_{35}=\exp(\alpha(e_{35}+e_{53}))$.

The Boyer-Lindquist-like coordinate is convenient for calculating the
geodesic orbits of test particles in twisted Kerr metric background.
It is well known that the geodesic equations are described by the
following Hamiltonian:
\bea
H(x, \pi ):=\frac{1}{2}g^{\mu \nu}p_{\mu }p_{\nu}
=\frac{1}{2}g^{\mu \nu}(\pi-eA)_{\mu}(\pi-eA)_{\nu}, \nonumber
\eea
where $(x, \pi)$ are canonical coordinates and
$\displaystyle p^{\mu}=\frac{dx^{\mu}}{d\lambda}=:\dot{x}^{\mu}$.
The $\lambda $ is the proper time coordinate.
The corresponding Hamilton-Jacobi equation
is written as
\bea
\frac{\partial  S}{\partial  \lambda }
+H\left( x, \frac{\partial  S}{\partial  x}\right) =0. \nonumber
\eea
The inverse metric is easily calculated and results in
\bea
g^{\mu \nu}\left( \frac{\partial }{\partial  x^{\mu }}\right) \otimes
\left( \frac{\partial }{\partial  x^{\nu}}\right)
&=&-\frac{1}{\Delta \rho^{2}}((R^{2}+a^{2})\partial_{t}
+a \partial_{\varphi})^{2}
+\frac{1}{\rho^{2} \sin^{2}\theta}
(\partial_{\varphi}+a\sin^{2}\theta \partial_{t})^{2} \nonumber\\
&&+\frac{\Delta}{\rho^{2}}(\partial_{r})^{2}
+\frac{1}{\rho^{2}}(\partial_{\theta})^{2}.
\label{invmetric}
\eea
We put the principal function $S(x)$ as
$\displaystyle S=\frac{1}{2}\mu ^{2}\lambda  -Et+L\varphi +S(r)
+s(\theta )$.
This means that the test particle under consideration has its square
of the norm of 4-velocity
$\mu^{2}$, electric charge $e$, angular momentum $L$, and energy
at infinity $E$.
The Hamilton-Jacobi equation is then
\bea
&&\mu ^{2}-\frac{1}{\Delta  \rho ^{2}}
(-(R^{2}+a^{2})E+aL-eQr)^{2}
+\frac{1}{\rho ^{2}\sin^{2}\theta }
(L-Ea\sin^{2}\theta)^{2} \nonumber \\
&&~~~~~~~~~~~~~~~~~~~~~~+\frac{\Delta }{\rho ^{2}}(S'(r))^{2}
+\frac{1}{\rho ^{2}}(s'(\theta))^{2}=0.
\label{hjeq}
\eea
Multiplying the factor $\rho ^{2}$, the variables are separated and
the equation of motion is reduced to two ordinary differential
equations with a separation constant $\nu$;
\bea
&&(S'(r))^{2}=\frac{\nu}{\Delta}+\frac{1}{\Delta^{2}}
[-(R^{2}+a^{2})E+aL-eQr]^{2}-\frac{R^{2}\mu^{2}}{\Delta},
\nonumber \\
&&(s'(\theta ))^{2}=-\nu -\frac{1}{\sin^{2}\theta}
[L-Ea\sin^{2}\theta]^{2}.
\nonumber
\eea
Geodesic motions are given by integrating the equations
$\displaystyle \dot{x}^{\mu }=p^{\mu }
=\left(  \frac{\partial S}{\partial x^{\nu}}\right)  g^{\mu \nu}$.
We will use this result when we discuss the thermodynamic property of
twisted Kerr black hole in section 5.
%
\setcounter{equation}{0}
\section{Kerr-Schild coordinate}
In this section, we construct the Kerr-Schild-like coordinate of the
twisted Kerr solution. It shows us that the twisted Kerr solution
can be analytically extended to the region of negative values of the
radial coordinate just as well as the Kerr or the Kerr-Newmann
solutions. We will do the extension by gluing two copies of the
metric
(\ref{ks}). It will also be shown that
singularity in the shape of a ring appears on the boundary of the
connection of these two copies.

We observed in the preceding section
that the twisting of the Kerr solution by $\Omega_{35}$
induces, roughly speaking, the replacement
$r^{2}\longrightarrow R^{2}=r^{2}+\beta r$.
this suggests us to modify the coordinate
transformation in the untwisted Kerr coordinate and we get to the
following coordinate transformation
$(t, \varphi, r, \theta)\longmapsto (\tau, x, y, z)$:
\bea
&&x:=(R\cos\tilde{\varphi}+a\sin\tilde{\varphi})\sin\theta, \;\;\;
y:=(R\sin\tilde{\varphi}-a\cos\tilde{\varphi})\sin\theta, \nonumber\\
&&z:=R\cos\theta, \;\;\;
d \tau := dt+dR-\frac{R^{2}+a^{2}}{\Delta}dR, \;\;\;
d\tilde {\varphi}:= d\varphi- \frac{a}{\Delta }dR.
\label{coordtrf}
\eea
Then the metric is rewritten as
\bea
ds^{2}&=&-d\tau^{2}+dx^{2}+dy^{2}+dz^{2}
-\frac{\beta^{2}}{4R^{2}+\beta^{2}}dR^{2}
\label{ks} \\
&&+\frac{(2m+\beta)R^{2}r}{R^{4}+a^{2}z^{2}}
\left[d\tau-\frac{zdz}{R}-\frac{R}{R^{2}+a^{2}}(xdx+ydy)
        -\frac{a}{R^{2}+a^{2}}(xdy-ydx)\right]^{2}.
\nonumber
\eea
The existence of the term
$\displaystyle -\frac{\beta^{2}}{4R^{2}+\beta^{2}}dR^{2} $
indicates the deviation from the standard Kerr-Schild form
$ds^2 =(\eta_{\mu\nu}+l_{\mu}l_{\nu})dx^{\mu}dx^{\nu}$.

Now, we investigate where the singularity lies in the twisted Kerr
spacetime. We follow the Chandrasekhar's calculation\cite{chandra}.
The metric of the
twisted Kerr solution can be rewritten as
\bea
ds^{2}=-e^{2\nu }dt^{2}
+e^{2\psi }(d\varphi-\omega dt)^{2}+e^{2p}dr^{2}
+e^{2q}d\theta ^{2}. \nonumber
\eea
Here
$\nu$, $\psi$, $\omega$, $p$, and $q$ are functions of $(r, \theta)$.
In the present case they are given by
\bea
&&e^{2\nu }=\frac{\rho ^{2}\Delta }{\Sigma ^{2}}, \;\;\;
e^{2\psi }=\frac{\Sigma ^{2}}{\rho ^{2}}\sin^{2}\theta , \;\;\;
\omega  =\frac{2Mar}{\Sigma ^{2}}, \;\;\;
e^{2p}=\frac{\rho ^{2}}{\Delta }, \;\;\;
e^{2q}=\rho ^{2}, \nonumber
\eea
together with
\bea
&&\Sigma ^{2}:=(R^{2}+a^{2})^{2}-\Delta  a^{2}\sin^{2}\theta , \;\;\;
\rho ^{2}:=R^{2}+a^{2}\cos^{2}\theta , \nonumber \\
&&M:=m+\beta /2, \;\;\;
\Delta :=r^{2}-2mr+a^{2},\;\;\;
R^{2}:=r^{2}+\beta r. \nonumber
\eea

After tedious calculations, we get explicit forms of $R_{abij}$'s.
Only information we need here is in which points these components blow
up. We show here only the result:
\it
\\

$\Sigma^{2}$ and $\rho^{2}$ are the only factors that can appear in
the denominators of components of $R_{abij}$. In other words,
curvature singularities appear only in points on which
$\Sigma^{2}$ or
$\rho^{2}$ (or both) becomes zero.
\rm
\\

Thus we have only to regard zeros of $\Sigma^{2}$ and $\rho^{2}$. We
provide a lemma to conclude that there is no singularity in the region
$r>0$:
\\
\\
{\bf Lemma}
$\;\;\;
\left\{ m\geq 0,\;\;\;\beta\geq 0,\;\;\; r>0\right\} \Longrightarrow
\Sigma^{2}>0,\;\;\; \rho^{2}>0. \;\;\; \frac{~~~~~~}{~~~~~~}$
\\
\\
Before we prove the lemma, we give a remark here. Clearly,
$\cosh\alpha -1\geq 0$ should hold, or $\alpha$ will not be real and
the reality of the electric charge $Q:=-m\sinh\alpha/\sqrt{2}$ will be
lost. That is, $\beta:=m(\cosh\alpha-1)$ and $m$ should be in the same
signature. Total mass (energy) of the system $M:=m+\beta/2$ should be
positive if the system describes a black hole at all. Therefore both
$m\geq 0$ and $\beta\geq 0$ are necessary.
The assumption $\left\{ \beta\geq 0,\;\; m\geq 0  \right\}$
in the lemma is quite reasonable in our case.
If we prove the lemma, we can say, under fairly general situations,
that there is no singularity in the
region $r>0$.
\\
\\
{\bf Proof of Lemma}\\
$\rho^{2}>0$ immediately follows from
$r>0$, and therefore $R^{2}=r^{2}+\beta r>0$ and
$a^{2}\cos^{2}\theta \geq 0$. We prove $\Sigma^{2}>0$ by dividing
the region $r>0$ into two: $\Delta \leq 0$ and $\Delta >0$.
If $\Delta \leq 0$, $\Sigma^{2}>0$ is trivial. If $\Delta >0$, then
\bea
\Sigma^{2}=(R^{2}+a^{2})^{2}-a^{2}\Delta  \sin^{2}\theta
\geq (R^{2}+a^{2})^{2}-a^{2}\Delta
=R^{4}+a^{2}r^{2}+2a^{2}\beta  r+2mra^{2}
\nonumber
\eea
holds. The right hand side of this inequality is strictly positive.
This completes the proof of the lemma. ~~~~~ \rule{4mm}{4mm} \\

Let us denote two copies of the spacetime by $A$ and $B$. Let $A$ be
the spacetime defined by $(R, \theta, t, \varphi)$ with
$R=+\left[
\frac{1}{2}(\delta +\sqrt{\delta ^{2}+4a^{2}z^{2}})
\right]^{1/2}>0$.
Similarly, let $B$ be the spacetime defined by
$(R, \theta, t, \varphi)$ with
$R=-\left[
\frac{1}{2}(\delta +\sqrt{\delta ^{2}+4a^{2}z^{2}})
\right]^{1/2}<0$.
We identify the two discs
$\left\{ x^{2}+y^{2}\leq  a^{2}, z=+0 \right\}$
in $A$ and
$\left\{ x^{2}+y^{2}\leq  a^{2}, z=-0 \right\}$
in $B$. And we do the same thing with $z=+0$ and $z=-0$ interchanged.
With these identifications, we obtain an analytically extended
solution, in which the radial coordinate $R$ is now defined in
$(-\infty, +\infty)$.
Singularity appears on $\left\{ \rho^2 =0 \cup \Sigma^2 =0 \right\}
=\left\{ R=0,\;\; \theta =\pi/2   \right\}$.
Obviously, it is in the shape of a ring and it locates on the boundary
of the two discs\cite{chandra}\cite{h-ellis}.
The global structure is the same as that of the untwisted Kerr.

To conclude, what we have acquired as knowledge of the twisted
Kerr solution is: \\
\\
\it
1) The twisting $\Omega_{35}$ gives the Kerr solution new
singularities. They appear, however, only in the region $r\leq 0$. \\
2) The analytically extended spacetime of the twisted Kerr solution
is so constructed that one removes the region $r<0$ out of the
spacetime ${\cal S}:=\{ (t,\varphi, r, \theta)  \}$ to obtain
the spacetime $\tilde{\cal S}:=\{ (t,\varphi, R, \theta)  \}$,
and one glues two copies of $\tilde{\cal S}$.
\rm
\\
\\
As we have mentioned at the end of section 2, $\Omega_{35}-$ twisting
of both the Schwarzschild and the Kerr solutions have new
singularities in the region $r\leq 0$. Despite of these singularities,
we can carry out the analytic extension of the twisted Kerr solution
by way of the transition of the radial coordinate $r$ into the new
one $R=\sqrt{r^2 +\beta r}$.
%
\setcounter{equation}{0}
\section{Thermal properties of the twisted Kerr black hole}
We calculate the Hawking temperature\cite{therm}
and other macroscopic quantities
of the twisted Kerr black hole via geodesic motions of test
bodies\cite{wald.ann}.
We will show that this calculation reproduces the thermodynamic
property of black holes\cite{wald}\cite{dilmass}.

Before we set out calculation, we draw a rough sketch. We consider
a test body with (mass=energy, electric charge, angular momentum)
$=(\delta M, \delta Q, \delta J)$,
thrown from infinity and going into the black hole of
mass=$M$, charge=$Q$, and angular momentum=$J$.
(Let us assume that $|\delta M| \ll |M|$, $|\delta Q| \ll |Q|$,
$|\delta J| \ll |J|$.)
When the test body goes into the black hole across the horizon,
physical quantities of the black hole will change:
\bea
(M, Q, J) \longrightarrow (M+\delta M, Q+\delta Q, J+\delta J).
\nonumber
\eea
We already know that if the electric charge or angular momentum
of the black hole exceed certain limit, horizon will disappear and
it will lead to a naked singularity\cite{sen}. Now, a question arises:
\lq\lq Can we change a singularity surrounded by regular
horizons into a naked one through the injection of test bodies?"
One has to add less $\delta M$ and more $\delta Q$ or $\delta J$ if
one wishes to change a black hole into a naked singularity. But,
there is obviously a lower limit of $\delta M$ for given $\delta Q$
and $\delta J$, because certain amount of energy is at least needed
in order to overcome the repulsive force made by the electric
interaction and the effective potential of angular momenta.
Unless $\delta M$ exceeds the lower limit $(\delta M)_{min}$,
the test body will not get to the black hole horizon and will move
back to infinity.

We estimate the $(\delta M)_{min}$ for given $\delta Q$
and $\delta J$ in the twisted Kerr metric background, and we will see
that it is impossible to change a black hole into a naked singularity
by way of the injection of test bodies. Furthermore, the inequality
of $\delta M$, $\delta Q$, and $\delta J$ can be considered as a
differential system in the $(M, Q, J)$-space and it can be integrated
to obtain the entropy and temperature of the black hole.

First, $(M, Q, J)$ of the solution (\ref{senmetric})
is expressed by the parameters $(m, a, \beta)$:
\bea
M=m+\frac{\beta}{2}, \;\;\; J=a(m+\frac{\beta}{2})=aM,
\;\;\; Q=\frac{m\sinh \alpha}{\sqrt{2}}. \nonumber
\eea
Its inverse is given by
\bea
m=M-\frac{Q^{2}}{2M}, \;\;\; a=\frac{J}{M}, \;\;\;
\beta =\frac{Q^{2}}{M}. \nonumber
\eea

As we saw in section 3, the orbit of the test particle
under consideration is expressed by the principal function
\bea
S(x)=\frac{1}{2}\mu^{2}\lambda -t\delta M+\varphi \delta J
+S(r)+s(\theta).
\nonumber
\eea
Clearly, the minimum for $\delta M$ is achieved when
$\displaystyle \mu=\frac{d\theta}{d\lambda}=0$. Hamilton-Jacobi
equation (\ref{hjeq}) is now given by
\bea
-\frac{1}{\Delta\rho^{2}}
[-(R^{2}+a^{2})\delta M+a\delta J+Qr\delta Q]^{2}
+\frac{1}{\rho^{2}\sin^{2}\theta}
(\delta J-\delta M a\sin^{2}\theta)^{2}
+\frac{\Delta}{\rho^{2}} (S'(r))^{2}=0 \nonumber
\eea
The minimum energy $(\delta M)_{min}$ is evaluated by the condition
in which $\displaystyle \frac{dr}{d\lambda}=g^{rr}S'(r)=0$
on the horizon $r=r_{+}:=m+\sqrt{m^2 -a^2}$.
That is,
\bea
-(R^{2}+a^{2})|_{r=r_{+}} (\delta M)_{min}+a\delta J+Qr_{+}\delta Q=0.
\nonumber
\eea
This means
\bea
\delta M \geq  \frac{a\delta J+Qr_{+}\delta Q }{2Mr_{+}}
=\frac{a\delta J}{2Mr_{+}}+\frac{Q\delta Q}{2M}.
\label{ineq}
\eea

Let us check that it is really impossible to
change a black hole into a naked singularity.
It suffices to check this when the black hole is
extremal: $m^2 =a^2$, i.e., $\displaystyle M^2 =(Q^2 /2)+J$.
Then $r_+ =J/M$. Putting this equation into (\ref{ineq}),
we get
\bea
\delta M\geq \frac{\delta J}{2M}+\frac{Q\delta Q}{2M}.
\nonumber
\eea
This is equivalent to the differential of $M^2 \geq Q^2 /2 +J$,
which is exactly the condition of the existence of the regular
horizon:
\bea
\delta (M^2 \geq Q^2 /2 +J) \Longleftrightarrow
2M\delta M \geq Q\delta Q+\delta J.
\nonumber
\eea
Therefore, even if one starts with the extremal black hole,
one cannot change $(M, Q, J)$ in such a way as to
break the inequality $M^2 \geq Q^2 /2 +J$.

Now, we show that the inequality (\ref{ineq}) can be
\lq\lq integrated" and it reproduces the thermal property of
the black hole. Let us consider the differential system
\bea
dM -\frac{Q}{2M}dQ-\frac{J}{2M^{2}r_{+}}dJ=0
\label{diffsys}
\eea
in the 3-dimensional space ${\bf R}^3$ parametrised by $(M, Q, J)$.
That is,
there is given a surface element which is normal to
$dM-(Q/2M)dQ-(J/2M^2 r_+ )dJ$ at each point in ${\bf R}^3$.
One can solve (\ref{diffsys}) by multiplying a suitable factor
on the both sides of (\ref{diffsys}) so that it satisfies the
integrability condition. We choose as the factor
\bea
2M+(2M^{2}-Q^{2})\left[
\left( M-\frac{Q^{2}}{2M} \right)^{2}-\left( \frac{J}{M} \right)^{2}
\right]^{-1/2}.
\label{factor}
\eea
Then (\ref{diffsys}) is integrated and we obtain
\bea
d
\left\{ M
   \left[ M-\frac{Q^{2}}{2M}
      +\sqrt{ \left( M-\frac{Q^{2}}{2M} \right)^{2}
        -\left( \frac{J}{M} \right)^{2} } \;\;
   \right]
\right\} =  0.
\nonumber
\eea
This means that the 2-dimensional surface with the defining equation
\bea
M
   \left[ M-\frac{Q^{2}}{2M}
      +\sqrt{ \left( M-\frac{Q^{2}}{2M} \right)^{2}
        -\left( \frac{J}{M} \right)^{2} } \;\;
   \right]
=  \mbox{const.}
\nonumber
\eea
in ${\bf R}^3$ is a solution of (\ref{diffsys}).
Noticing that the factor (\ref{factor}) is positive,
the inequality (\ref{ineq}) is rewritten as
\bea
\delta
\left\{ M
   \left[ M-\frac{Q^{2}}{2M}
      +\sqrt{ \left( M-\frac{Q^{2}}{2M} \right)^{2}
        -\left( \frac{J}{M} \right)^{2} } \;\;
   \right]
\right\} \geq  0.
\nonumber
\eea
This inequality is precisely equivalent to
\bea
\delta
\left\{
\frac{1}{8\pi}(\mbox{Area of the horizon})
\right\}
\geq 0.
\nonumber
\eea
Actually,
\bea
\mbox{Area}
&=&\int_{r=r_{+}} \sqrt{g_{\theta \theta }g_{\varphi\varphi}}
d\theta \wedge d\varphi
=4\pi \left.(R^{2}+a^{2})\right|_{r=r_{+}} =4\pi (2m+\beta )r_{+}
\nonumber \\
&=&8\pi M
\left[ M-\frac{Q^{2}}{2M}
+\sqrt{ \left( M-\frac{Q^{2}}{2M}\right)^{2}
-\left( \frac{J}{M} \right)^{2} } \;\;
\right] .
\nonumber
\eea
Thus, the gedanken experiments of throwing test particles into
the twisted Kerr black hole really reproduce the 2nd law of
black hole mechanics.

Now that the entropy $S$ ($1/4$ of the area of the horizon)
of the black hole is given in terms of $(M, Q, J)$,
one can calculate its differential
\bea
dS&=&\left(\frac{\partial  S}{\partial  M}\right)dM
+\left(\frac{\partial  S}{\partial  J}\right)dJ
+\left(\frac{\partial  S}{\partial  Q}\right)dQ.
\nonumber
\eea
Comparing this with the formula of the 1st law of the black hole
mechanics
$dM=TdS+\Phi dQ+\Omega dJ$, one obtains
\bea
T=\frac{\sqrt{\#}}
        { 4\pi M \left( M-\frac{Q^{2}}{2M}+\sqrt{\#} \right)},
\;\;\;
\Omega =\frac{J}{2M^{2}\left( M-\frac{Q^{2}}{2M}+\sqrt{\#}\right)},
\;\;\;
\Phi =\frac{Q}{2M},
\label{phyq}
\eea
where we put $\displaystyle \#:=\left( M-\frac{Q^{2}}{2M}\right)^{2}
-\left( \frac{J}{M} \right)^{2}$.

On the other hand, surface gravity $\kappa$, angular velocity
$\Omega_H$, and electrostatic potential $\Phi_H$ on the horizon
are given by
\bea
&&\frac{1}{4}\kappa :=
\left.\sqrt{\frac{-1}{2}\nabla^{\mu}l^{\nu}\nabla_{\mu}l_{\nu}}\right|_{H}
=\lim_{r \rightarrow r_{+}} \sqrt{g^{rr}}
\partial _{r}\sqrt{-g_{tt}}|_{\theta=0}
=\frac{\sqrt{\#}}
        { 2M \left( M-\frac{Q^{2}}{2M}+\sqrt{\#} \right)},\nonumber\\
&&\Omega _{H}
:=-\left[ \frac{g_{t\varphi}}{g_{\varphi\varphi}}\right]_{H}
=\frac{J}{2M^{2}r_{+}}, \nonumber \\
&&\Phi_{H} :=\left[l^{\mu }A_{\mu }\right]_{H}
=\frac{Q}{2M} \nonumber
\eea
where $\displaystyle l^{\mu }:=
\left( \frac{\partial }{\partial  t}
\right)^{\mu }+\Omega _{H}
\left( \frac{\partial }{\partial  \varphi}
\right)^{\mu }$.
These are exactly the same as (\ref{phyq}).
The gedanken experiments of test
particles are again consistent with the 1st law of black hole
mechanics.
%
\setcounter{equation}{0}
\section{$O(2,3)$ symmetry and asymptotic flatness}
We have seen that the action (\ref{haction}) admits the global
$O(2,3)$ symmetry and that the vacuum Kerr metric is a solution of
the equations of motion. One might hope that one can produce a larger
class of solutions by acting elements of $O(2,3)$. However, we
immediately realise that not all of the elements of $O(2,3)$
preserve the asymptotic flatness. And some elements cause merely
coordinate transformations and they do not change physical contents.
Here a question arises: \lq\lq how many
physically distinct black hole solutions are generated by the $O(2,3)$
transformation?". We cannot answer this question completely at
present, but part of the information can be cast from the
infinitesimal $o(2,3)$ action, which we are going to discuss.

We will show in this section that
$\left\{ \exp (\alpha_{35}(e_{35}+e_{53}))
\left| \alpha_{35}\in {\bf R} \right.\right\} \subset O(2,3)$
is the only one-parameter subgroup that preserves the
asymptotic flatness.
And all the other elements of $O(2,3)$ break the asymptotic flatness
or cause only coordinate transformations.
%
\subsection{$o(2,3)$ and $gl(2,{\bf R})$ transformations}
Let us explain the basic strategy.
As we have seen in section 2,
the action of $O(2,3)$ mixes $\hat{g}_{mn}$, $\hat{A}_{m}$, and
$\hat{B}_{mn}$ ($(t, \varphi)-$ components) each other
nonlinearly and reproduces their new configuration.
On the other hand, $\tilde{A}_{\alpha }$ and $\tilde{B}_{\alpha \beta}$ are
left
invariant and $\tilde{g}_{\alpha \beta}$ ($(r, \theta )-$ components) are
only multiplied by a function
$\mbox{det}(\Omega(\hat{g}))/\mbox{det}(\hat{g})$.
So, if the twisting by an element of $O(2,3)$ breaks the
asymptotic behaviour and yet the breaking is \lq\lq soft" enough to
be recovered by a coordinate transformation of $(t, \varphi)$, then
one will obtain a favourable new solution.
\[ \begin{array}{ccccc}
\left( \begin{array}{c} \mbox{seed} \\ \mbox{solution} \end{array}
\right)
& \spmapright{O(2,3)} &
\left( \begin{array}{c} \mbox{***} \\ \mbox{***} \end{array}
\right)
& \spmapright{ GL(2,{\bf R}) }&
\left( \begin{array}{c} \mbox{asymptotically flat} \\ \mbox{solutions}
        \end{array}
\right)
\end{array} \]
Here $GL(2, {\bf R})$ denotes the affine transformation group of
$(t, \varphi)$. Since we are considering stationary and axisymmetric
solutions, $GL(2, {\bf R})$ should be a global symmetry. We linearise these
transformations to seek well-behaved twisted solutions.

First, we check the asymptotic behaviour.
When the spacetime is asymptotically flat, the fields behave as
\bea
&&g_{tt}=-1+\frac{2M}{r}+O(r^{-2}),\;\;\;
g_{t\varphi}=\frac{-4J}{r}\sin^2 \theta +O(r^{-2}), \nonumber\\
&&g_{\varphi\varphi}=r^2 \sin^2 \theta + O(r),\;\;\;
g_{rr}=1+O(r^{-1}), \;\;\;
g_{\theta\theta}=r^2 + O(r), \nonumber\\
&&A_t =\frac{Q}{r}+O(r^{-2}), \;\;\;
A_{\varphi}=O(r^{-1}),\;\;\;
B=O(r^{-1}),\;\;\;
\phi= O(r^{-1}).
\nonumber
\eea
And the corresponding behaviour of the matrix-valued field
${\cal M}(r, \theta)$ is given by
\bea
{\cal M} =
\left( \matrix{
-4 & 0  &0&0&0 \cr 0& p+2 &0&p&0 \cr 0&  0  &0&0&0 \cr
0&p&0& p-2 &0 \cr 0&0&0&0&0 \cr }
\right) +O(r^{-1});
\nonumber
\eea
where $p:=p_{2}r^{2}+p_{1}r+p_{0}$ and
\bea
p_2 :=\lim_{r \rightarrow \infty}r^{-2}g_{\varphi\varphi},\;\;\;
p_1 :=\lim_{r \rightarrow \infty}r^{-1}(g_{\varphi\varphi}-p_2 r^2 ), \;\;\;
p_0 :=\lim_{r \rightarrow \infty} (g_{\varphi\varphi}-p_2 r^2-p_1 r).
\nonumber
\eea
That is, $p$ is the divergent and constant parts of
$g_{\varphi\varphi}$ as the radial coordinate $r$ goes to infinity.
We assume that $r$ should remain to be the radial coordinate
before and after
the twisting.
We will go back and discuss this assumption at the end
of this section.

We consider a general element of $\Omega \in O(2,3)$ and $P \in
GL(2,{\bf R})$, and act $P \cdot \Omega $ on a seed solution. We
examine its linearisation $1\cdot d \Omega +dP \cdot 1$.
Now, let us act elements of $o(2,3)$ and $gl(2,{\bf R})$ and check the
asymptotic behaviour of infinitesimally twisted solutions.
Clearly, $o(2,3)$ is spanned by the following canonical basis:
\bea
o(2,3)=
\left\{ \matrix{
e_{12}+e_{21},&
e_{13}+e_{31},&
-e_{14}+e_{41},\cr
-e_{15}+e_{51},&
-e_{23}+e_{32},&
e_{24}+e_{42},\cr
e_{25}+e_{52},&
e_{34}+e_{43},&
e_{35}+e_{53},\cr
-e_{45}+e_{54}&&\cr
}\right\}_{\bf R}
=:\left\{
d\Omega_{ij}
\left|
1\leq i < j \leq 5
\right.
\right\}_{\bf R} ,
\nonumber
\eea
where $e_{ij}$'s are matrix units of $M_{5}({\bf R})$. We recall
that the action of $O(2,3)$ on \\
${\cal M}(\hat{g}_{\mu\nu}, \hat{B}_{\mu\nu}, \hat{A}_{\mu})$ is given by
$\Omega ({\cal M})=\Omega  {\cal M}\Omega '$.
Therefore its differential is
\bea
d\Omega ({\cal M})=d\Omega {\cal M}+{\cal M}d\Omega'
=:d
\left(\matrix{
l&n&u \cr
n'&c&e \cr
u'&e'& f \cr
}\right). \nonumber
\eea
We choose $d\Omega\in o(2,3)$ as the most general form:
\bea
d\Omega =\sum_{1\leq i<j\leq 5}\alpha_{ij}d\Omega_{ij} \in o(2,3)
\;\;\;\;(\forall \alpha_{ij}\in {\bf R}).
\nonumber
\eea
The result of the calculation of $d\Omega({\cal M})$ is given by
\bea
&&(d\Omega({\cal M}))_{11}=O(r^{-1}), \;\;\;
(d\Omega({\cal M}))_{12}=-2\alpha_{12}+(\alpha_{12}-\alpha_{14})p
+O(r^{-1}), \nonumber \\
&&(d\Omega({\cal M}))_{13}=-4\alpha_{13}+O(r^{-1}), \;\;\;
(d\Omega({\cal M}))_{14}=-2\alpha_{14}+(\alpha_{12}-\alpha_{14})p
+O(r^{-1}) \nonumber \\
&&(d\Omega({\cal M}))_{15}=-4\alpha_{15}+O(r^{-1}), \;\;\;
(d\Omega({\cal M}))_{22}=2\alpha_{24}p+O(r^{-1}),  \nonumber \\
&&(d\Omega({\cal M}))_{23}=2\alpha_{23}+(\alpha_{23}+\alpha_{34})p
+O(r^{-1}), \;\;\;
(d\Omega({\cal M}))_{24}=2\alpha_{24}+O(r^{-1}),  \nonumber \\
&&(d\Omega({\cal M}))_{25}=2\alpha_{25}+(\alpha_{25}+\alpha_{45})p
+O(r^{-1}), \;\;\;
(d\Omega({\cal M}))_{33}=O(r^{-1}),  \nonumber \\
&&(d\Omega({\cal M}))_{34}=-2\alpha_{34}+(\alpha_{23}+\alpha_{34})p
+O(r^{-1}), \;\;\;
(d\Omega({\cal M}))_{35}=O(r^{-1}), \nonumber\\
&&(d\Omega({\cal M}))_{44}=2\alpha_{24}p+O(r^{-1}), \;\;\;
(d\Omega({\cal M}))_{45}=-2\alpha_{45}+(\alpha_{25}+\alpha_{45})p
+O(r^{-1}), \nonumber\\
&&(d\Omega({\cal M}))_{55}=O(r^{-1}).  \label{domegam}
\eea

The $GL(2,{\bf R})-$ action on the coordinate of $(t, \varphi)$
\bea
\left( \matrix{t \cr \varphi \cr }\right)
\longmapsto
P^{-1}\cdot
\left( \matrix{t \cr \varphi \cr }\right), \;\;\;\;\;
P
\in GL(2,{\bf R})
\nonumber
\eea
induces $P(\hat{g})=P\hat{g} P'$. Its differential is
$dP(\hat{g})=dP\hat{g}+\hat{g}dP'$, or equivalently,
$dP(\hat{g}^{-1})=-dP'\hat{g}^{-1}-\hat{g}^{-1}dP$ for
$dP \in gl(2,{\bf R})$.
We choose it in a general form as
$\displaystyle dP=\left( \matrix{ \beta_{11}&\beta_{12}\cr
\beta_{21}&\beta_{22}\cr} \right)$.

We immediately notice that $\alpha_{35}-$ component does not appear
in $d\Omega ({\cal M})$, whereas all the other components do.
We see from this that $\Omega_{35}$ is the only element of $O(2,3)$
that preserves the asymptotic flatness. But there is still hope that
the breaking of the asymptotic condition is just superficial and the
asymptotic flatness can be reinstated by some coordinate
transformations. We will see in the next subsection, however,
that all of such \lq\lq soft" twisting only result in coordinate
transformations.
\subsection{detailed calculations}
In this subsection, we first calculate
$d\hat{g}=(d\Omega +dP)(\hat{g})$ and $d\hat{B}=(d\Omega +dP)(\hat{B})$
to obtain necessary conditions to keep the asymptotic behaviour.
These conditions will tell us that only suitable linear combinations
of $(d\Omega_{ij})$ are allowed. And then we will explicitly calculate
the allowed transformations.
It will reveal that all the allowed transformations except $\Omega_{35}$
are equivalent to coordinate transformations and they give no
physically new solutions.

First, we check the asymptotic behaviour of
$d\hat{g}=(d\Omega +dP)(\hat{g})$ and $d\hat{B}=(d\Omega +dP)(\hat{B})$.
{}From eq.(\ref{m2gab}), we have
$\displaystyle d\hat{g}^{-1}
=\frac{1}{4}\eta (dl-dn-dn'+dc)\eta $.
This enables us to calculate $d\hat{g}$ explicitly:
$(d\Omega+dP)(\hat{g})=-\hat{g}( d\Omega (\hat{g}^{-1}))\hat{g}
+dP \hat{g}+\hat{g} dP$.
The result is:
\bea
d\hat{g}_{tt}&=&-2\alpha _{13}-2\beta _{11}+O(r^{-1}) \nonumber \\
d\hat{g}_{t\varphi}&=&
-\frac{1}{2}(\alpha _{12}+\alpha _{14}-\alpha _{23}+\alpha _{34})
-\beta _{21}
+\frac{1}{2}p(\alpha _{12}-\alpha _{14}+\alpha _{23}+\alpha _{34})
+p\beta _{12}+O(r^{-1}) \nonumber \\
d\hat{g}_{\varphi\varphi}&=&2\alpha _{24}p+2\beta _{22}p+O(r^{-1}).
\nonumber
\eea
We get
\bea
&&\alpha _{13}=-\beta _{11}, \;\;\;
\alpha _{12}+\alpha _{14}-\alpha _{23}+\alpha _{34}+2\beta _{21}=0,
\nonumber \\
&&\alpha _{12}-\alpha _{14}+\alpha _{23}+\alpha _{34}+2\beta _{12}=0,
\;\;\;
\alpha _{24}=-\beta _{22}
\label{dgcond}
\eea
as necessary conditions in order to maintain the asymptotic
behaviour of $\hat{g}_{mn}$.

After similar but longer calculations, we get the result of $d\hat{B}$:
\bea
d\hat{B}_{t\varphi}=\frac{1}{2}(-\alpha _{12}-\alpha _{14}-\alpha _{23}+\alpha
_{34})
-\frac{1}{2}p(-\alpha _{12}+\alpha _{14}+\alpha _{23}+\alpha _{34})
+O(r^{-1}).
\nonumber
\eea
This gives
\bea
-\alpha _{12}-\alpha _{14}-\alpha _{23}+\alpha _{34}=0,
\;\;\;
-\alpha _{12}+\alpha _{14}+\alpha _{23}+\alpha _{34}=0
\label{dbcond}
\eea
as necessary conditions.

Now, let us pay attention to the components of
$(d\Omega +dP)({\cal M})_{i5}$.
Noticing that
$\displaystyle \lim_{r \rightarrow \infty}{\cal M}_{i5}=0$,
$dP({\cal M})_{i5}=O(r^{-1})$ holds.
This means that the constant terms and
divergent terms which appear in
$d\Omega ({\cal M})_{i5}$ cannot be cured
by $dP({\cal M})$. Thus
\bea
\alpha_{15}=\alpha_{25}=\alpha_{45}=0
\label{dacond}
\eea
is necessary in view of the asymptotic behaviour.

Gathering all the data of (\ref{dgcond}), (\ref{dbcond}), and
(\ref{dacond}), we have
\bea
&&\alpha _{12}=\alpha _{34}=-\frac{1}{2}(\beta _{12}+\beta _{21}),
\;\;\;
\alpha _{14}=-\alpha _{23}=\frac{1}{2}(\beta _{12}-\beta _{21}),
\nonumber \\
&&\alpha_{15}=\alpha_{25}=\alpha_{45}=0, \;\;\;
\alpha _{13}=-\beta _{11},\;\;\; \alpha _{24}=-\beta _{22},
\;\;\; \alpha _{35}:\;
\mbox{free}.
\label{data}
\eea
This shows that the only 5 elements of
\bea
\left\{
d\Omega_{12}+d\Omega_{34}, \;\;\;
d\Omega_{14}-d\Omega_{23}, \;\;\;
d\Omega_{13}, \;\;\;
d\Omega_{24}, \;\;\;
d\Omega_{35}
\right\}
\label{5genes}
\eea
are allowed generators.
All the other 5 elements break the asymptotic flatness so severely
that the breaking cannot be recovered
by coordinate transformations.

Next, we examine the twisting generated by the exponentials of
the above elements (\ref{5genes}).
After tedious and boring calculations, one finds that the
twisting by $\exp (\alpha (d\Omega_{12}+d\Omega_{34}))$
is exactly equivalent to the coordinate transformation
$$(t, \varphi)\longmapsto (t\cosh \alpha -\varphi \sinh \alpha , -t \sinh\alpha
+\varphi \cosh\alpha ),$$
which is naturally expected from (\ref{data}).
The twisting by
$\exp (\alpha (d\Omega_{14}-d\Omega_{23}))$,
$\exp (\alpha  d\Omega_{13})$, and $\exp(\alpha  d\Omega_{24})$ , all of them
result in coordinate transformations as well.
The equivalence is listed as:
\bea
\left.
\matrix{GL(2, {\bf R})&&O(2,3) \cr
\exp(-\alpha  (e_{12}+e_{21}))& \Longleftrightarrow &
\exp(\alpha (d\Omega_{12}+d\Omega_{34})) \cr
\exp(-\alpha e_{11}) &\Longleftrightarrow &
\exp(\alpha d\Omega_{13}) \cr
\exp(-\alpha  (-e_{12}+e_{21}))& \Longleftrightarrow &
\exp(\alpha (d\Omega_{14}-d\Omega_{23})) \cr
\exp(-\alpha e_{22})& \Longleftrightarrow &
\exp(\alpha d\Omega_{24}). \cr
}
\right.
\label{embed}
\eea
Notice that the map $gl(2, {\bf R}) \hookrightarrow o(2,3)$
\bea
&&e_{11}\longmapsto d\Omega_{13}, \;\;\;
e_{12}\longmapsto
\frac{1}{2}(d\Omega_{14}-d\Omega_{23})
-\frac{1}{2}(d\Omega_{12}+d\Omega_{34}), \nonumber \\
&&e_{22}\longmapsto d\Omega_{24}, \;\;\;
e_{21}\longmapsto
-\frac{1}{2}(d\Omega_{14}-d\Omega_{23})
-\frac{1}{2}(d\Omega_{12}+d\Omega_{34})
\nonumber
\eea
gives the (Lie-algebraic) embedding of
$gl(2, {\bf R})$ in $o(2,3)$ and it is in tune with the
correspondence (\ref{embed}).

Thus we are left with only one twisting
of $\exp (\alpha d\Omega_{35})$, which is already well known to us.
\subsection{physical interpretation}
We conclude this section by some remarks and discussion.
The calculations in the last subsection gives us the information
that
\it
\lq\lq $\Omega_{35}$ is the only one-parameter subgroup of $O(2,3)$
that transforms a seed solution into another and at the same time
keeps stationarity, axisymmetry, and asymptotic flatness."
\rm

Actually, suppose that one twists a seed solution by the element
$\Omega \in O(2,3)$ in the most general form:
$\displaystyle \Omega=\Omega (\alpha )
=\exp (\sum_{i<j} \alpha _{ij} d\Omega_{ij})$.
Arguments of the infinitesimal transformation has shown that
one must impose
$\displaystyle \sum_{i<j} \alpha _{ij} d\Omega_{ij}$ be in
5 dimensional subspace of $o(2,3)$ spanned by (\ref{5genes}).
Now, if one restricts the domain of $(\alpha _{ij})$ to a suitably
small neighbourhood of $0 \in {\bf R}^{10}$, an arbitrary element
of the form
$\displaystyle \exp (\sum_{i<j} \alpha _{ij} d\Omega_{ij})$
is expressed as
\bea
\exp(\alpha _{12}'d\Omega_{12})
\cdot\exp(\alpha _{13}'d\Omega_{13})
\cdots
\exp(\alpha _{45}'d\Omega_{45}).
\nonumber
\eea
(This is just the diffeomorphism between the normal coordinates
of the 1st and 2nd kinds.) Therefore one can suppose without
loss of generality that $\Omega (\alpha )$ is in the form of
\bea
\exp(\alpha _{1}(d\Omega_{12}+d\Omega_{34}))
\cdot\exp(\alpha _{2}d\Omega_{13})
\cdot\exp(\alpha _{3}(d\Omega_{14}-d\Omega_{23}))
\cdot\exp(\alpha _{4}d\Omega_{24})
\cdot\exp(\alpha _{5}d\Omega_{35}).
\nonumber
\eea
Since one already knows that
\bea
\exp(\alpha _{1}(d\Omega_{12}+d\Omega_{34}))
\cdot\exp(\alpha _{2}d\Omega_{13})
\cdot\exp(\alpha _{3}(d\Omega_{14}-d\Omega_{23}))
\cdot\exp(\alpha _{4}d\Omega_{24})
\nonumber
\eea
induces only a coordinate transformation, $\Omega(\alpha )-$
twisting gives
a solution which is equivalent to the $\Omega_{35}-$ twisting.

It should be noted that our arguments given here are developed
in the vicinity of the identity of $O(2,3)$. Both the linearisation
and the diffeomorphism between the normal coordinates of the 1st
and 2nd kinds are \lq\lq local" arguments.
So, one cannot deny the
possibility of constructing a new solution by way of multiple twists.
That is, even though each of the twists breaks the asymptotic
condition, they might form a nice transformation as a whole.
For example, one can perhaps construct a new solution in such a way
that one breaks the asymptotic flatness by twisting
by a certain element of $\Omega_{1} \in O(2,3)$, and then transforms
it along the flow of $\{ \exp(\alpha \omega) | \alpha  \in {\bf R},\;\;
\omega \in o(2,3)   \}$, and finally twists it by a suitable element
of $\Omega_{2} \in O(2,3)$ so that the resulting solution can
satisfy the asymptotic flatness:
\bea
\Omega_{2}\cdot \exp(\alpha \omega) \cdot \Omega_{1}
(\mbox{seed solution}).
\nonumber
\eea
We did not consider such cases. The readers will agree that
the investigation of the \lq\lq global" property of the $O(2,3)$
transformation is far more difficult than that of the local one.
We will put off this problem for future investigation.

Finally, let us discuss the assumption of the radial coordinate $r$.
We have assumed that $r$ should remain to be a radial coordinate
before and after the twisting.
One might wonder that we have put too strict assumption on $r$. That
is, one might imagine that even if there is some harmful divergence
of the field components at the spatial infinity $r=\infty$,
it could be amended by suitable coordinate redefinitions of the
radial coordinate. Such a miracle, however, seldom happens.
Now, let us consider the question \lq\lq can one recover
the asymptotic flatness by introducing a new radial
coordinate $\xi=\xi(r)$ ?".
As we have seen in the previous subsection, all the infinitesimal
transformations of $\alpha_{ij}d\Omega_{ij} \in o(2,3)$,
except for $\alpha_{13}d\Omega_{13}$,
$\alpha_{24}d\Omega_{24}$, and $\alpha_{35}d\Omega_{35}$,
add terms with positive powers of $r$ to $g_{t\varphi}$ or
$\hat{A}_{m}$. Then these fields will diverge both at $r=\infty$
and $r=0$, since they contain terms with both negative and positive
power of $r$. One will have to give up
identifying the spatial
infinity as $r=\infty$ or as $r=0$.
Therefore the new radial coordinate $\xi$,
if it exists, should satisfy
\bea
\left\{ \xi =\infty \right\}
=\left\{
        r=\mbox{(a certain point $q$, which is not
        $0$ nor $\infty$)}
\right\}.
\nonumber
\eea
On the other hand, asymptotic behaviour demands that
\bea
&&\Omega(g)_{\xi\xi}(q)=
e^{-2\Omega(\phi)(q)}g_{rr}(q)\left( \frac{dr}{d\xi}\right)^2 (q)
\sim 1, \nonumber\\
&&\Omega(g)_{\theta\theta}(q)=
e^{-2\Omega(\phi)(q)}g_{\theta\theta}(q) \sim \xi^2,
\nonumber\\
&&\Omega(g)_{\varphi\varphi} \sim \xi^2 \sin^2 \theta,
\nonumber\\
&& \Omega (A_t )(q) \sim 0, \;\;\; \Omega (A_\varphi )(q)\sim 0,\;\;\;
\Omega B_{t\varphi})(q) \sim 0.
\nonumber
\eea
It is quite difficult to find a new radial coordinate that satisfies
such an asymptotic condition.
%
\setcounter{equation}{0}
\section{Summary and conclusion}
We analysed the twisted Kerr solution.
The operation of the twisting
$\Omega_{35}=\exp(\alpha d\Omega_{35})$ does
not drastically change the global structure of the spacetime.
The Kerr-Schild-like coordinate is
introduced to show where the singularities
lie in the spacetime.
With a suitable radial coordinate transformation $r \mapsto R
=\sqrt{r^2 + \beta r}$, we can analytically extend the twisted
Kerr solution just as we did in the (untwisted) Kerr solution.
The twisting generates new singularities
in the region $r<0$. But the transition $r \mapsto R$ naturally
cuts away the negative
radial part $r<0$.
Analytic continuation is carried out by gluing the two charts of
$R\geq 0$ and $R\leq 0$.

In section 5, we considered null geodesics of the twisted Kerr
metric and discussed its thermal behaviour.
The gedanken experiments of throwing test particles into
the black hole have led us to the consistent
description of the thermodynamic-like property of the black hole.
That is, \\
1) Test bodies can cross the horizon and go into the
black hole if and only if the condition (\ref{ineq})
is satisfied. We cannot decrease the area of the horizon by throwing
test bodies into the black hole.\\
2) We cannot change the black hole into a naked singularity
by any means of the injection
of test bodies.\\
3) Physical quantities of the black hole $(\kappa, \Omega_H , Q_H )$
are reproduced by \lq\lq integrating" the condition (\ref{ineq}).

We have neglected the interaction of the dilaton and the antisymmetric
tensor with particles. We considered only null geodesics,
so the interaction of the dilaton can be removed\cite{campbell}.
Interaction of the antisymmetric tensor is more subtle. We put this
problem for future investigation.

We discussed in section 6 the global $O(2,3)$ symmetry and
asymptotic flatness.
Some elements of
$O(2,3)$ break the asymptotic flatness and some elements cause only
coordinate transformations.
Infinitesimal $o(2,3)$ action reveals that 5 generators of $o(2,3)$
break the asymptotic condition so severely that the breaking cannot
be cured by coordinate transformations. And thus we are left with
5 generators listed in (\ref{5genes}). All the generators except for
$d\Omega_{35}$ cause coordinate transformations and they do not
change physics. Therefore we cannot add any more free parameters
than mass, angular momentum, and electric charge, because the
asymptotically flat, stationary, and axisymmetric solution we
already know is the $\Omega_{35}-$ twisted Kerr solution only.
This gives quite an affirmative example of the no-hair
conjecture\cite{nohair} of stationary and axisymmetric stringy
black hole solutions.
\\
\\
\\
\footnotesize {\it Acknowledgments.}
I wish to express my gratitude to Prof. T.Eguchi for
many helpful discussions and suggestions. I acknowledge Prof.
J.Arafune for continuous encouragement and valuable comments.
I thank Prof. A.Kato and all the members of the Elementary
Particle Theory Group in the University of Tokyo for support
and collaboration.
\normalsize
%
\\

\end{document}